\documentclass[aps,pra,twocolumn,groupedaddress,nofootinbib,floatfix]{revtex4}
\par
\usepackage{graphicx} 
\usepackage{graphics}
\usepackage{epsfig}
\usepackage{braket}
\usepackage{bm}
\usepackage{amssymb}
\usepackage{mathtools}
\usepackage{amsmath}
\usepackage{float}
\begin{document}
	\title{Boosting Linear-Optical Bell Measurement Success Probability with Pre-Detection Squeezing and Imperfect Photon-Number-Resolving Detectors}

	\author{Thomas Kilmer and Saikat Guha}
	%
	%
	\affiliation{College of Optical Sciences, University of Arizona, Tucson, Arizona 85721, USA}

	\date{\today}
	
	\begin{abstract}
		Linear optical realizations of Bell state measurement (BSM) on two single-photon qubits succeed with probability $p_s$ no higher than $0.5$. However pre-detection quadrature squeezing, i.e., quantum noise limited phase sensitive amplification, in the usual linear-optical BSM circuit, can yield ${p_s \approx 0.643}$. The ability to achieve $p_s > 0.5$ has been found to be critical in resource-efficient realizations of linear optical quantum computing and all-photonic quantum repeaters. Yet, the aforesaid value of $p_s > 0.5$ is not known to be the maximum achievable using squeezing, thereby leaving it open whether close-to-$100\%$ efficient BSM might be achievable using squeezing as a resource. In this paper, we report new insights on why squeezing-enhanced BSM achieves $p_s > 0.5$. Using this, we show that the previously-reported ${p_s \approx 0.643}$ at single-mode squeezing strength $r=0.6585$---for unambiguous state discrimination (USD) of all four Bell states---is an experimentally unachievable point result, which drops to $p_s \approx 0.59$ with the slightest change in $r$. We however show that squeezing-induced boosting of $p_s$ with USD operation is still possible over a continuous range of $r$, with an experimentally achievable maximum occurring at $r=0.5774$, achieving ${p_s \approx 0.596}$. Finally, deviating from USD operation, we explore a trade-space between $p_s$, the probability with which the BSM circuit declares a ``success", versus the probability of error $p_e$, the probability of an input Bell state being erroneously identified given the circuit declares a success. Since quantum error correction could correct for some $p_e > 0$, this tradeoff may enable better quantum repeater designs by potentially increasing the entanglement generation rates with $p_s$ exceeding what is possible with traditionally-studied USD operation of BSMs.
	\end{abstract}

	\maketitle
	
	\section{Introduction}
	\label{sec:Introduction}
	
	In the so-called {\em dual-rail} photonic qubit basis, the logical qubits $|{\boldsymbol 0}\rangle$ and $|{\boldsymbol 1}\rangle$ are realized by one photon in one of two orthogonal modes (while the other mode is in vacuum). Two common physical realizations are: (1)~one photon in one of two orthogonal polarization modes (of a single spatio-temporal mode), i.e., ${|{\boldsymbol 0}\rangle = |1\rangle_{\rm H}}$ and ${|{\boldsymbol 1}\rangle = |1\rangle_{\rm V}}$; and (2)~one photon in one of two orthogonal spatial modes (of the same polarization), i.e., ${|{\boldsymbol 0}\rangle = |10\rangle}$ and ${|{\boldsymbol 1}\rangle = |01\rangle}$. In this paper we will consider the latter form of the dual-rail photonic qubit basis.
	
	The four Bell states form a complete orthonormal basis for two qubits. Each is a maximally entangled state of two qubits, representing $1$ ebit of shared entanglement. Equations~(\ref{eq:Bell States}) are the four $2$-qubit $4$-mode Bell states in the dual rail encoding, where the mode pairs 1\&2 and 3\&4 each represent one qubit. 
	\begin{subequations}
		\label{eq:Bell States}
		\begin{eqnarray}
		|\psi_+\rangle &=& \frac{1}{\sqrt{2}}(|1001\rangle+|0110\rangle)
		\\
		|\psi_-\rangle &=& \frac{1}{\sqrt{2}}(|1001\rangle-|0110\rangle)
		\\
		|\phi_+\rangle &=& \frac{1}{\sqrt{2}}(|1010\rangle+|0101\rangle)
		\\
		|\phi_-\rangle &=& \frac{1}{\sqrt{2}}(|1010\rangle-|0101\rangle)
		\end{eqnarray}
	\end{subequations}
	
	\par
	
	Projecting a pair of qubits onto the Bell basis is an important primitive in quantum computation and communication. This two-qubit projective measurement is often termed the Bell basis measurement, or the Bell state measurement (BSM). Uses of BSMs include quantum teleportation~\cite{Teleportation1,Teleportation2}, entanglement swapping~\cite{Swapping}, dense coding~\cite{Dense}, quantum repeaters~\cite{Communication1, Communication2}, and fault-tolerant quantum computing~\cite{Computation,Teleportation2}. 
	
	It has been long known that if the optical circuit used to realize a BSM on two dual rail qubits is restricted to using passive linear optical elements (beamsplitters and phase shifters), the BSM can succeed with at most $50\%$ success probability~\cite{50}. This probabilistic nature of the linear optical BSM has arguably been the primary bottleneck in realizing scalable all-photonic quantum computing~\cite{Computation,Percolation1} and other all-photonic quantum information processing, such as all-photonic quantum repeaters for long-distance entanglement distribution~\cite{Orep_Lo,Orep_Pant}, where an entangled cluster state of several photons in a `code state' mimics the action of a quantum memory by providing error correction against photon loss.
	
	\subsection{Boosted Bell State Measurements}
	
	The success probability~($p_s$) of a BSM circuit is the probability with which it can unambiguously identify a randomly assigned Bell state presented to the circuit as input. When restricted to passive linear optics and no ancillary states, one cannot achieve ${p_s>50\%}$~\cite{50}. However it was recently shown that including pre-detection quadrature squeezing and photon number resolving (PNR) detection (along with linear optics) can boost (increase) $p_s$ to $\sim 64.3\%$~\cite{64.3}. Further, it was also recently found that the addition of unentangled single-photon ancillae and PNR detection (along with linear optics) can boost $p_s$ to up to $78.125\%$~\cite{75}. The addition of arbitrarily large entangled ancillae and PNR detection (along with linear optics) can boost $p_s$ to 100\%~\cite{100}. 
	
	
	Any quantum measurement on a set of $n$ linear optical qubits can be realized by a suitable $2n$-mode quantum unitary transformation followed by ideal PNR detectors. A {\em universal} set of optical transformations is one from which one can draw elements to build an arbitrary unitary transformation on any set of optical modes. One example of a universal set comprises the following subsets: (1) passive linear optical elements (circuits involving beam splitters and phase shifters) and on-off detectors~\cite{LOuniversal}, (2) squeezing (realized, e.g., using an optical parametrical amplifier (OPA) built using a $\chi^{(2)}$ non-linear-optical material), and (3) PNR detectors. The most general multi-mode {\em Gaussian} transformation is one that can be built by drawing elements from subsets (1) and (2)~\cite{UniversalGaussian}. Adding any one non-Gaussian element---which can be either (a) a non-Gaussian non-linear {\em transformation} such as the Kerr or cubic phase, (b) a non-Gaussian ancilla {\em state} such as single photons or cat states or (c) a non-Gaussian {\em measurement} such as PNR detection---to the set of Gaussian transformations makes it a universal set~\cite{CVuniversal}. The linear optical circuit for BSM that achieves $p_s = 50\%$ (shown in Fig.~\ref{fig:Passive}) involves elements only from the subset (1) above. In fact, one cannot exceed this success probability with a $2n$-mode circuit comprised of elements drawn from (1) alone~\cite{50}. 
	
	Since optical elements from (1), (2) and (3) above form a universal set, a BSM circuit must exist that uses linear optics, squeezing, and PNR detectors that achieves $p_s = 100\%$. The boosted BSM circuit proposed by Zaidi and van Loock~\cite{64.3} that achieves $p_s \approx 64.3\%$ employs exactly those elements (linear optics, squeezing, and PNR detectors). This circuit is shown in Fig.~\ref{fig:Squeezed}. The universality of the elements used in realizing this circuit, and the practical realizability of inline squeezing and PNR detection~\cite{TES} are the reasons we will study boosted BSM using pre-detection squeezing in further detail, in this paper.
	
	\subsection{Main results}
	
	The main results of this paper are summarized below.
	
	\begin{enumerate}
		\item We review the squeezing-boosted BSM circuit in Ref.~\cite{64.3} and provide a deeper understanding and intuition for why it boosts $p_s$ beyond the $50\%$ limit of linear optics without ancillae.
		
		\item We show that the ${p_s \approx 64.3\%}$ result of~\cite{64.3} is actually a point result achievable with a pre-detection squeezing of {\em exactly} $5.719$~dB (squeezing parameter, ${r = 0.6585}$) in all four squeezers in the circuit in Fig.~\ref{fig:Squeezed}. If the applied squeezing in any of the squeezers is even infinitesimally lower or higher than that, $p_s$ drops (discontinuously) to a lower value, but it stays strictly higher than $50\%$ (see Fig.~\ref{fig:4 Rails, Squeezed, USD}).
		
		\item We evaluate the effect of finite photon number resolution in the PNR detectors. We find that: (1) at least $3$-photon resolving detectors are needed to get any boost of $p_s$ above $50\%$, specifically $p_s \approx 0.54$ being achievable with $3$-photon resolving PNR detectors and $3$ dB of squeezing ($r \approx 0.35$); (2) Most of the  non-discontinuous boost in $p_s$ possible from pre-detection squeezing, $\approx 59\%$, can be obtained with $7$-photon resolving PNR detectors (e.g., recently-demonstrated transition edge sensor (TES) detectors by NIST~\cite{TES}).
		
		\item In all of the existing research on boosted linear optic BSMs~\cite{50,64.3,75,100}, the definition of {\em success} of a BSM is based on the so-called unambiguous state discrimination (USD) framework, where one demands that when one of the four Bell states is presented to the BSM circuit at random, if the circuit does produce a result identifying the input as one of the four Bell states, it is $100\%$ certain that particular Bell state was input. With probability $1-p_s$, it produces an ``I don't know" (erasure) outcome. 
		
		With imperfect (sub-unity detection efficiency, non-zero dark click rate) PNR detectors, even the BSM circuits in Refs.~\cite{50,64.3,75,100} will {\em not} produce a USD outcome and cannot be analyzed as such. Further, in certain applications, it may be advantageous to trade a higher `success' probability $p_s$, the probability the circuit produces some non-erasure result, with a non-zero `error' probability ($p_e$), the probability that the produced measurement result is incorrect (for USD, $p_e = 0$).
		
		In this paper we study the performance of pre-detection-squeezing boosted BSM in such a non-USD, probabilistic state discrimination (PSD) framework, and work out the trade-space between $p_s$ and $p_e$. We also incorporate non-ideal (sub-unity efficiency) PNR detection into this analysis. 
	\end{enumerate}
	\section{Passive Linear Optics}
	\label{sec:50}
	\begin{figure}[H]
		\centering
		\includegraphics[width=3in]{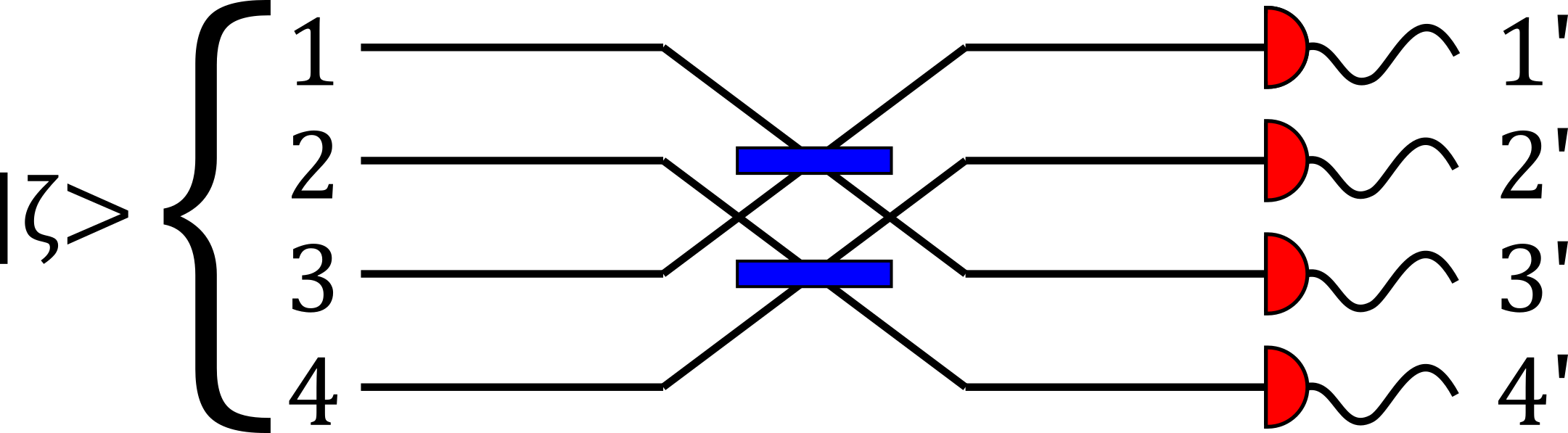}
		\caption{\label{fig:Passive} The optical circuit that identifies a randomly input Bell state $|\zeta\rangle\in\left\{|\psi_+\rangle,|\psi_-\rangle,|\phi_+\rangle,|\phi_-\rangle\right\}$ with success probability $p_s = 0.5$. With probability $1 - p_s = 0.5$, it produces an erasure (``I don't know") outcome. This is an example of unambiguous state discrimination (USD).}
	\end{figure} 
	
	The four Bell states~(\ref{eq:Bell States}) are mutually orthogonal quantum states, and hence perfectly and deterministically distinguishable by an appropriate quantum measurement. However, if all four modes of a (randomly) presented Bell state are detected using on-off detectors or PNR detectors, the Bell states are perfectly {\em indistinguishable}. This is because none of the four possible photon number detection patterns ${\left(N_1,N_2,N_3,N_4\right) \in \left\{1001, 0110, 1010, 0101\right\}}$ can uniquely identify the input Bell state, where $N_i$ is the number of photons detected in mode $i$. The four $4$-mode product number states ${\left\{|1001\rangle, |0110\rangle, |1010\rangle, |0101\rangle\right\}}$ are each present in two Bell states. So the observation of a given number state does not uniquely identify a particular Bell state. Therefore the USD BSM success probability achieved by direct PNR detection is ${p_s=0\%}$. 
	
	However the use of beam splitters prior to detecting the four modes, as shown in Fig.~\ref{fig:Passive} can improve $p_s$ to the $50\%$ limit achievable with passive linear optics~\cite{50}. In this paper we will define a `beam splitter' to always refer to the following 50-50 beam splitter~(\ref{eq:Beam Splitter}):
	\begin{equation}
	\begin{pmatrix} {a_k^\prime}^\dag \\ {a_l^\prime}^\dag \end{pmatrix} = 
	\frac{1}{\sqrt{2}}\begin{pmatrix} 1 & i \\ i & 1 \end{pmatrix} 
	\begin{pmatrix} a_k^\dag \\ a_l^\dag \end{pmatrix}.
	\label{eq:Beam Splitter}
	\end{equation}
	Mixing modes 1\&3 and 2\&4 of the input Bell state on two beam splitters as shown in Fig.~\ref{fig:Passive}, produces the following output states:
	\begin{subequations}
		\label{eq:50:50 Bell States}
		\begin{eqnarray}
		|\psi^\prime_+\rangle  &=&  \frac{i}{\sqrt{2}}(|1100\rangle+|0011\rangle),
		\\
		|\psi^\prime_-\rangle &=& \frac{1}{\sqrt{2}}(|1001\rangle-|0110\rangle),
		\\
		|\phi^\prime_+\rangle &=& \frac{i}{2}(|2000\rangle+|0020\rangle+|0200\rangle+|0002\rangle),
		\\
		|\phi^\prime_-\rangle &=& \frac{i}{2}(|2000\rangle+|0020\rangle-|0200\rangle-|0002\rangle).
		\end{eqnarray}
	\end{subequations}
	The four output states in~(\ref{eq:50:50 Bell States}) are still mutually orthogonal and perfectly distinguishable as were the original input Bell states, since the two beamsplitters comprise a unitary (and hence reversible) transformation. However, with on-off detection or PNR detection on these output states, ${p_s = 0.5}$. In order to see why, note that $|\psi^\prime_+\rangle$ and $|\psi^\prime_-\rangle$ are completely distinguishable with photon detection, while $|\phi^\prime_+\rangle$ and $|\phi^\prime_-\rangle$ remain wholly indistinguishable. Therefore, if one of the four Bell states is presented randomly (i.e., each with probability $1/4$), the circuit in Fig.~\ref{fig:Passive} will unambiguously identify the input state with average probability $50\%$. This circuit provides the framework for all subsequent {\em boosted} BSM circuits we will examine in this paper. 
	\section{Pre-Detection Squeezing}
	\label{sec:squeezing}
	\begin{figure}[H]
		\centering
		\includegraphics[width=3in]{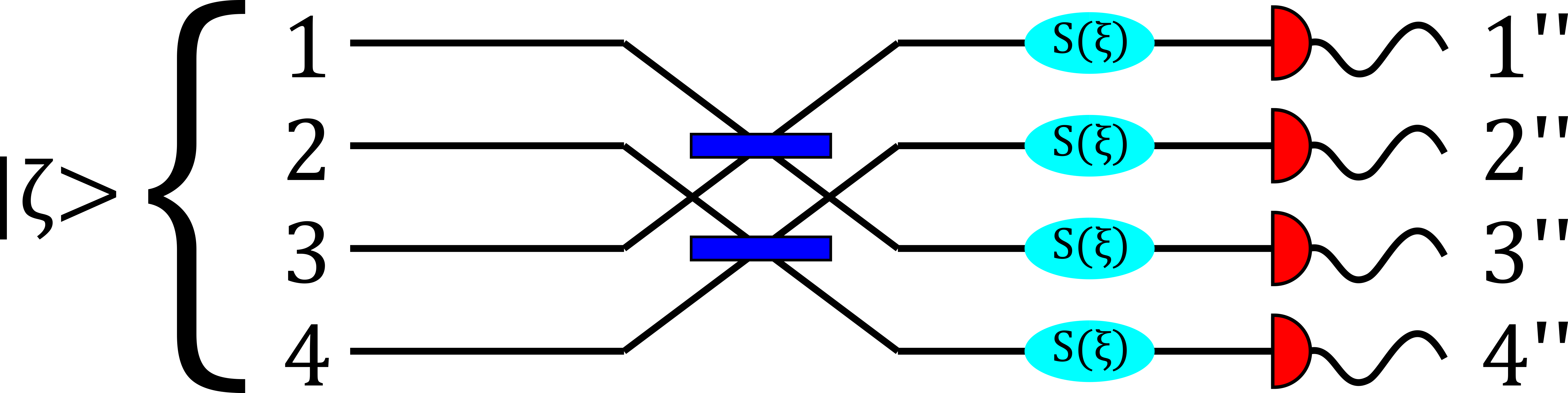}
		\caption{\label{fig:Squeezed}An optical circuit which can identify unambiguously a randomly input Bell state $|\zeta\rangle\in\left\{|\psi_+\rangle,|\psi_-\rangle,|\phi_+\rangle,|\phi_-\rangle\right\}$ with $p_s > 50\%$. $S(\xi)$ represents a single-mode squeezer (e.g., an OPA), where $\xi = \frac12 e^{i\phi} {\rm tanh} r$ is the degree of squeezing.}
	\end{figure} 
	Zaidi and van Loock's BSM circuit functions by applying pre-detection single-mode quadrature squeezing to each of the four output modes of the standard $p_s = 50\%$ BSM circuit~\cite{64.3}, as shown in Figure~\ref{fig:Squeezed}. The action of single-mode squeezing $S(\xi)$, a unitary (hence, reversible) operation, on the number state $|n\rangle$ is given by~\cite{Snum}:
	\begin{subequations}
		\label{eq:Squeezed Number States}
		\begin{eqnarray}
		S(\xi)|n\rangle &=& \left(\frac{1}{\text{cosh}(2r)}\right)^{n+1/2}\left(n!\right)^{1/2}
		\\
		&&\sum_{j=0}^{\left[\frac{n}{2}\right]}\frac{\left(-\xi^*\right)^j\left(\text{cosh}(2r)\right)^{2j}}{\left(n-2j\right)!j!} 
		\nonumber \\
		&&\sum_{k=0}^{\infty}\frac{\xi^k\left[\left(n-2j+2k\right)!\right]^{1/2}}{k!}|n-2j+2k\rangle
		\nonumber \\
		\xi &\equiv& \frac{1}{2}e^{i\phi}\text{tanh}~r
		\\
		\text{dB} &=& -10 \log_{10}\left(e^{-2r}\right)
		\end{eqnarray}
	\end{subequations}
	
	Assuming identical squeezing intensity $r$ and rotation $\phi=0$ for each mode, the output states of the circuit in Fig.~\ref{fig:Squeezed} for the four input Bell states are given in Eqs.~\eqref{eq:Squeezed Bell States}. Note that these double-primed transformed Bell states are squeezing applied on the single-primed transformed states~\eqref{eq:50:50 Bell States} at the output of the standard linear-optical BSM circuit in Fig.~\ref{fig:Passive}. Appendix~\ref{ap:Non-Uniform} goes into greater detail about why we choose $r$ and $\phi$ to be the same for all four modes.
	
	Let us note from Eq.~\eqref{eq:Squeezed Number States} that squeezing is a parity conserving operation, which maps a number state $|n\rangle$ to an infinite superposition of all number states of like parity, i.e., if $n$ is odd (resp., even), $S(\xi)|n\rangle$ is a superposition of all odd (resp., even) number states. Additionally, let us define the parity of a set of modes in a product number state as the list of parities of the number of photons in each of those modes. For example, the parities of the (four-mode) number state components in $|\psi^\prime_+\rangle$ are \mbox{(odd, odd, even, even)} and \mbox{(even, even, odd, odd)}, whereas the parity of the number state components in $|\psi^\prime_-\rangle$ are \mbox{(odd, even, even, odd)} and \mbox{(even, odd, odd, even)}. 
	\newpage{\pagestyle{empty}\cleardoublepage}
	~
	\begin{widetext}
		\begin{subequations}
			\label{eq:Squeezed  Bell States} 
			\begin{eqnarray}
			|\psi^{\prime\prime}_+(r)\rangle  &=&  \frac{i\left(\text{sech}~r\right)^4}{\sqrt{2}}|1100\rangle + \frac{i\left(\text{sech}~r\right)^4}{\sqrt{2}}|0011\rangle + \frac{i\left(\text{sech}~r\right)^4\text{tanh}~r}{2}|0211\rangle + \frac{i\left(\text{sech}~r\right)^4\text{tanh}~r}{2}|2011\rangle
			\\
			&&+ \frac{i\left(\text{sech}~r\right)^4\text{tanh}~r}{2}|1120\rangle + \frac{i\left(\text{sech}~r\right)^4\text{tanh}~r}{2}|1102\rangle + \frac{i\left(\text{sech}~r\right)^4\left(\text{tanh}~r\right)^2}{2}|1122\rangle + \ldots
			\nonumber \\
			\nonumber \\
			|\psi^{\prime\prime}_-(r)\rangle  &=&  \frac{\left(\text{sech}~r\right)^4}{\sqrt{2}}|0110\rangle - \frac{\left(\text{sech}~r\right)^4}{\sqrt{2}}|1001\rangle + \frac{\left(\text{sech}~r\right)^4\text{tanh}~r}{2}|0112\rangle + \frac{\left(\text{sech}~r\right)^4\text{tanh}~r}{2}|2110\rangle
			\\
			&&- \frac{\left(\text{sech}~r\right)^4\text{tanh}~r}{2}|1021\rangle - \frac{\left(\text{sech}~r\right)^4\text{tanh}~r}{2}|1201\rangle + \frac{\left(\text{sech}~r\right)^4\left(\text{tanh}~r\right)^2}{2\sqrt{2}}|2112\rangle + \ldots
			\nonumber \\ 
			\nonumber \\
			|\phi^{\prime\prime}_+(r)\rangle &=& -i\sqrt{2}\left(\text{sech}~r\right)^2\text{tanh}~r|0000\rangle - \frac{i\left(\text{sech}~r\right)^4\left(\text{cosh}(2r)-2\right)}{2}|2000\rangle - \frac{i\left(\text{sech}~r\right)^4\left(\text{cosh}(2r)-2\right)}{2}|0200\rangle
			\\
			&&- \frac{i\left(\text{sech}~r\right)^4\left(\text{cosh}(2r)-2\right)}{2}|0020\rangle - \frac{i\left(\text{sech}~r\right)^4\left(\text{cosh}(2r)-2\right)}{2}|0002\rangle + \ldots
			\nonumber \\ 
			\nonumber \\
			|\phi^{\prime\prime}_-(r)\rangle &=& \frac{i\left(\text{sech}~r\right)^4}{2}|2000\rangle - \frac{i\left(\text{sech}~r\right)^4}{2}|0200\rangle + \frac{i\left(\text{sech}~r\right)^4}{2}|0020\rangle - \frac{i\left(\text{sech}~r\right)^4}{2}|0002\rangle
			\\
			&&+ \frac{i\left(\text{sech}~r\right)^4\text{tanh}~r}{\sqrt{2}}|2020\rangle - \frac{i\left(\text{sech}~r\right)^4\text{tanh}~r}{\sqrt{2}}|0202\rangle + \frac{i\left(\text{sech}~r\right)^4\left(\text{tanh}~r\right)^2}{4}|2220\rangle + \ldots
			\nonumber
			\end{eqnarray}
		\end{subequations}
	\end{widetext} 
	
	Since single-mode squeezing on a number state is parity preserving, squeezing on a set of modes in a product number state is also parity preserving (by our definition of parity of a set of modes). Therefore, squeezing has no effect on the distinguishability of number states products that have unique parity. For example, the parities of the states of $|\psi^\prime_+\rangle$ and $|\psi^\prime_-\rangle$ remain unchanged after applying squeezing, and hence $|\psi^{\prime\prime}_+\rangle$ and $|\psi^{\prime\prime}_-\rangle$ remain perfectly distinguishable by PNR detection on all four modes regardless of the amount of squeezing applied. However number states with like parity signatures, such as those of $|\phi^\prime_+\rangle$ and $|\phi^\prime_-\rangle$, the parity of all the number states within which is {(even,~even,~even,~even)}, are mapped to an overlapping set of number states, each with {(even,~even,~even,~even)} parity.
	
	Even though $|\phi^\prime_+\rangle$ and $|\phi^\prime_-\rangle$ were completely indistinguishable by PNR detection, some distinguishability emerges between the post-squeezed states $|\phi^{\prime\prime}_+\rangle$ and $|\phi^{\prime\prime}_-\rangle$. There are two distinct ways in which such distinguishability comes about: 
	\begin{enumerate}
		\item For all values of $r$, certain new number states, e.g., ${\left\{|0000\rangle, |2002\rangle, |0220\rangle,\text{ etc.}\right\}}$, appear in $|\phi^{\prime\prime}_+\rangle$ but not in $|\phi^{\prime\prime}_-\rangle$, and vice versa. The detection of one of these states by PNR detection thus boosts $p_s$ beyond $50\%$, while maintaining USD distinguishability. The probability of occurrence of these distinguishing states, and thus the boost in $p_s$, is a function of $r$ (see Fig.~\ref{fig:4 Rails, Squeezed, USD}).
		\item For certain discrete values of $r$, one or more of the states ${\left\{|2000\rangle,~|0200\rangle,~|0020\rangle,~|0002\rangle,\text{ etc.}\right\}}$, which $|\phi^{\prime\prime}_+\rangle$ and $|\phi^{\prime\prime}_-\rangle$ are both comprised of, interfere completely destructively for  $|\phi^{\prime\prime}_+\rangle$ but not for $|\phi^{\prime\prime}_-\rangle$, or vice versa. This produces additional USD distinguishability. However, this happens only for certain singular values of $r$ and thus produces discontinuous boosts to $p_s$ (see the two black dots in Fig.~\ref{fig:4 Rails, Squeezed, USD}).
	\end{enumerate}
	
	We performed a numerical calculation of the average distinguishability between the four Bell states, the details of which are given in Subappendix~\ref{subsec:USD}. The result, a plot of BSM success probability $p_s$ as a function of squeezing intensity $r$ (squeezing phase $\phi = 0$) is shown in Fig.~\ref{fig:4 Rails, Squeezed, USD}.  
	\begin{figure}[!h!t]
		\centering
		\includegraphics[width=\columnwidth]{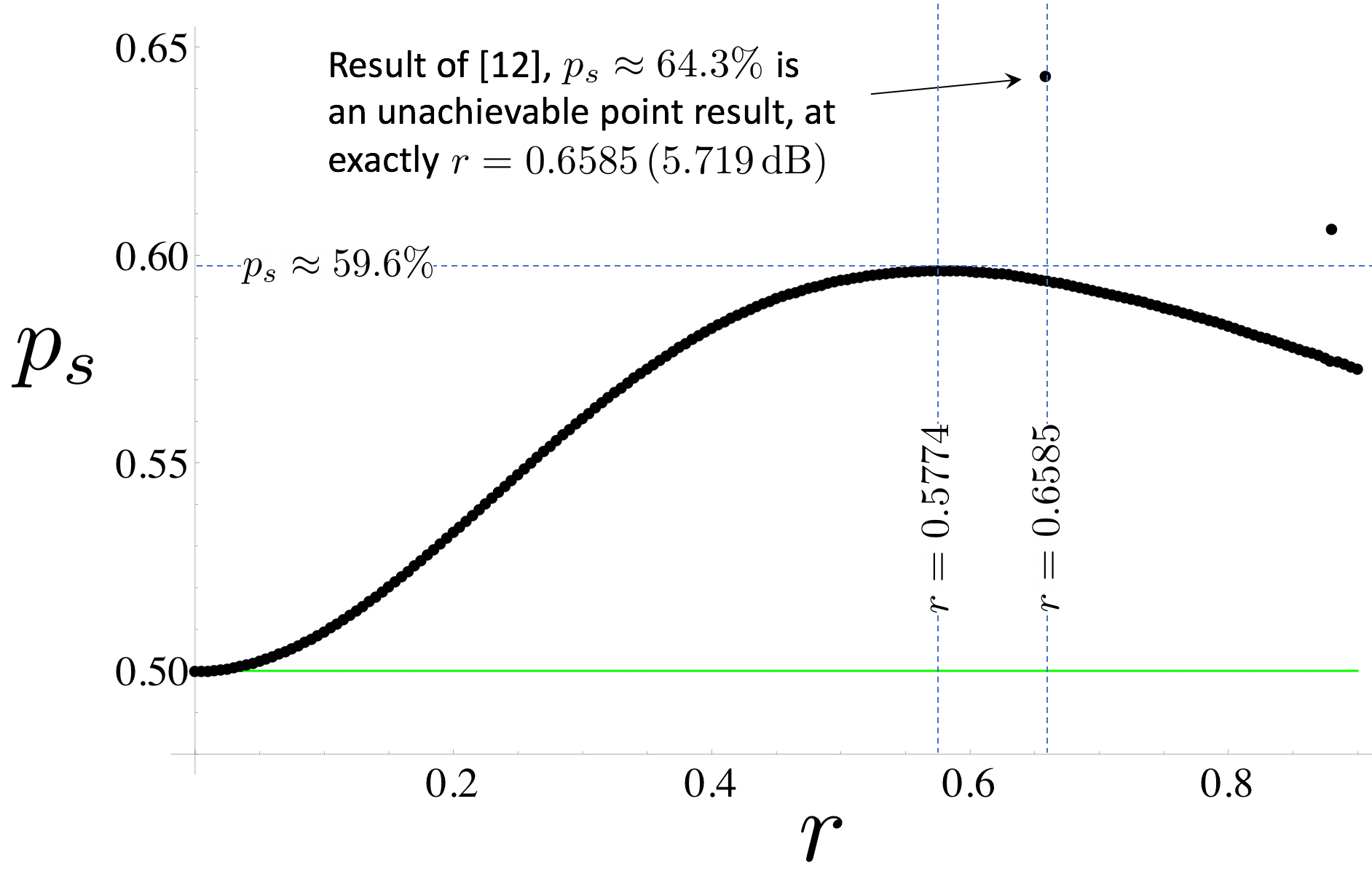}
		\caption{\label{fig:4 Rails, Squeezed, USD}Bell measurement success probability ($p_s$) vs. squeezing intensity ($r$) for a squeezed Bell measurement circuit.
		}
	\end{figure} 
	
	Ref.~\cite{64.3} reported the measurement success peak at $r=0.6585$ ($5.719$ dB of squeezing), where ${p_s=64.3\%}$. This peak occurs when ${\text{cosh}(2r)-2=0}$, causing the ${|2000\rangle,~|0200\rangle,~|0020\rangle,\text{ and }|0002\rangle}$ number states to destructively interfere within $|\phi^{\prime\prime}_+\rangle$, making them unique to $|\phi^{\prime\prime}_-\rangle$. This adds an additional $4.94\%$ success probability. There are infinitely many such discontinuous peaks in $p_s$ at different values of $r$ but this particular peak is the only one worth considering, as it occurs at the lowest $r$, has the highest relative increase in $p_s$, and has the highest absolute $p_s$ of any such peak. 
	
	Unfortunately these peaks only exist for infinitesimally thin slices of $r$. They are discontinuous jumps in success probability and are not reasonable targets for USD measurements, as any uncertainty in $r$ will cause the number states in question to become indistinguishable with PNR detection. Further attention will be given to these peaks later, when considering PSD measurements. 
	
	However, despite those discontinuous peaks, what is interesting is that along the entire continuous portion of the $p_s$ vs. $r$ curve in Fig.~\ref{fig:4 Rails, Squeezed, USD}, for every value of $r$, USD BSM is possible at $p_s > 50\%$. This contribution to the continuous boost in $p_s$ comes from the squeezed number states that are perfectly distinguishable for all values of $r$. Considering only these universally distinguishable states we find a maximum success probability of ${p_s=59.6\%}$ at $r=0.5774$ ($5.015$ dB of squeezing). 

	\subsection{Finite Photon Number Resolution}
	\label{subsec:Finite Photon Number Resolution}
	
	Squeezing a number state produces an infinite sum (superposition) of number states with arbitrarily large photon numbers. However existing state of the art PNR detectors, the so-called transition edge sensor (TES) detectors, cannot efficiently resolve number states with more than $7$ photons~\cite{TES}. Using such detectors in the squeezing-boosted BSM circuit will truncate the number states beyond a certain number, which should be taken into consideration when evaluating the success probability. Figure~\ref{fig:4 Rails, Squeezed, USD, Limited} shows the $p_s$ curves for a squeezed BSM circuit with PNR detectors limited to a maximum photon number resolution ($n_{\rm max}$) of $12$ or fewer photons. These curves are compared to the $p_s$ curve with arbitrarily high photon number resolution.
	
	Reasonably sized arrays of on-off detectors can be used to emulate the performance of PNR detectors for small photon numbers~\cite{PNR_emulation}, which could be a more experimentally accessible realization of PNR detection than TES detectors for most laboratories. However in practice even small arrays of on-off detectors greatly increase the number of photon number states. The computational resources required to characterize such systems increase exponentially with array size. We believe such an analysis may not be infeasible, but is beyond our reach given the computational resources at our disposal. 
	\begin{figure}[!h!t]
		\centering
		\includegraphics[width=\columnwidth]{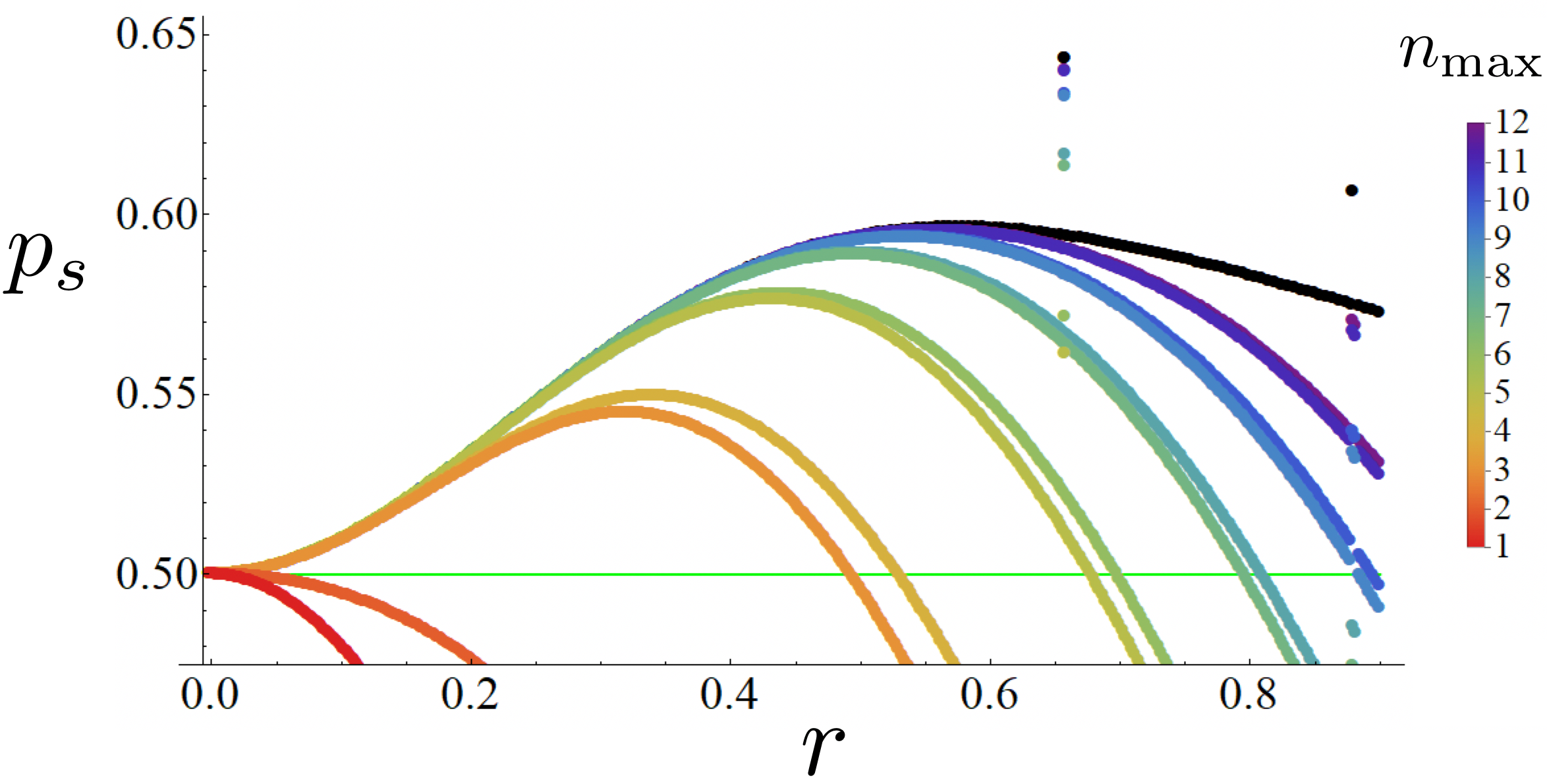}
		\caption{\label{fig:4 Rails, Squeezed, USD, Limited}Bell measurement success probability ($p_s$) vs. squeezing intensity ($r$) for a pre-detection-squeezed BSM circuit, and with PNR detectors with maximum number resolution $n_{\rm max} \in \left\{1, 2, \ldots, 12\right\}$. The case of $n_{\rm max} = \infty$ from Fig.~\ref{fig:4 Rails, Squeezed, USD} is also shown for comparison (top black curve).
		}
	\end{figure} 
	
	PNR detectors limited to ${n_{\rm max}=7}$ show a (discontinuous) success probability peak ${p_s = 61.3\%}$ at $r=0.6585$ ($5.719$ dB of squeezing) and a continuous maxima of ${p_s=58.9\%}$ at $r=0.496$ ($4.311$ dB of squeezing). 
	\subsection{Probabilistic State Discrimination}
	\label{subsec:Probabilistic State Discrimination}
	
	So far we have limited our analysis to unambiguous state discrimination (USD), where one demands that when one of the four Bell states is presented to the BSM circuit at random, if the circuit does produce a result identifying the input as one of the four Bell states, it is $100\%$ certain that particular Bell state was input. In this USD analysis, we had the BSM circuit announce a determination for an input Bell state only for those detected number states which occur uniquely in the output state for exactly one Bell state, whether for a single value of $r$ or for all $r$. This limited us to the regime where our probability of erroneous state identification ($p_e$) is $0$. Here we will expand our performance evaluation to include number states which occur in multiple Bell states, allowing ${p_e > 0}$ to produce a higher success probability~($p_s$). Certain applications may benefit from operating in this regime, where an outer layer of quantum error correction may correct for some errors, but gain from a higher rate of success. 
	
	We call this operational regime probabilistic state discrimination (PSD), and define the following probabilities:
	\begin{itemize}
		\item $p_s + p_e$: probability that the BSM circuit declares a `success' in identifying the input Bell state.
		\item $1 - (p_s + p_e)$: erasure probability; BSM circuit declares an ``I don't know" outcome.
		\item $\alpha \equiv p_s/(p_s + p_e)$: measurement confidence; conditioned on the event that the BSM circuit declares a success, $\alpha$ is the probability that the input Bell state was indeed correctly identified.
	\end{itemize}
	
	$p_s$ and $p_e$ will be taken to be averaged over the chosen input Bell state (each of the four assumed to occur equally likely) and over all PNR detection outcomes. We will now analyze the $(p_s, p_e)$ tradeoff, obtained by appropriately post-processing the PNR detector outputs.
	
	When a number state occurs in multiple output states, instead of rejecting that outcome as erasure, we can use the relative probabilities of its occurrence in each Bell state to decide which Bell state to guess as been input. For instance at $r=0.6$ the state $|2000\rangle$ occurs in $|\phi^{\prime\prime}_+(0.6)\rangle$ and $|\phi^{\prime\prime}_-(0.6)\rangle$ with net probabilities $0.230\%$ and $6.41\%$ respectively. Therefore if we detect a $(2,0,0,0)$ click pattern on the PNR detectors, we can decide with $\sim96.5\%$ certainty that the $|\phi_-\rangle$ Bell state was the BSM circuit's input. This increases the success probability~($p_s$) of our BSM circuit by $1.54\%$ when compared with USD operation at $r=0.6$, while introducing a $0.0561\%$ chance of an erroneous measurement~($p_e$).
	
	Considering different subsets of these multiply-occurring number states as valid outcomes, as discussed above, will yield different $p_s/p_e$ ratios. For our analysis we set a maximum allowable~$p_e$ and then include that subset of duplicate number states which provides an optimal $p_s$ enhancement for that $p_e$, while still considering all other duplicate number states as erasure. 
	
	The process for calculating $p_s$ and identifying optimal subsets of duplicate number states for a given $p_e$ is detailed in Subappendix~\ref{subsec:Probabilistic}. 
	\begin{figure}[!h!tp]
		\centering
		\includegraphics[width=3in]{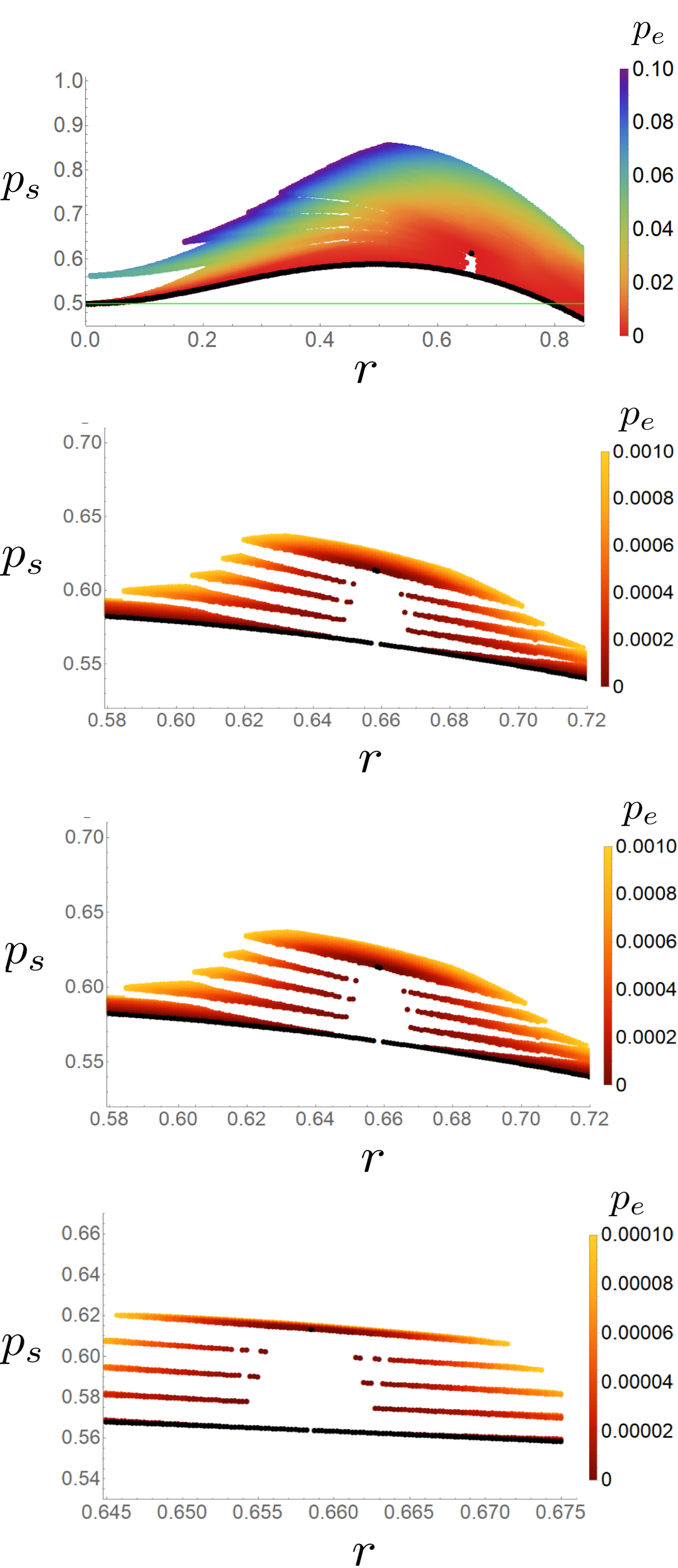}
		\caption{\label{fig:4 Rails, Squeezed, Prob, 7}Bell measurement success probability ($p_s$) vs. squeezing amplitude ($r$), with respect to error probability ($p_e$), for a pre-detection-squeezed BSM circuit, and with PNR detectors restricted to a maximum number resolution $n_{\rm max}=7$. The black dots represent the results of USD measurements with $p_e=0$. Error probability~($p_e$) is restricted to [0\%,10\%],~[0\%,1\%],~[0\%,0.1\%],~and~[0\%,0.01\%] in the top, second from top, second from bottom, and bottom plots respectively.}
	\end{figure}
	
	Figure~\ref{fig:4 Rails, Squeezed, Prob, 7} plots $p_s$ vs. $r$ for PSD operation, assuming lossless (unity detection efficiency) PNR detectors with ${n_{\rm max}=7}$ number resolution, for a range of error probabilities $p_e$. Note that the discontinuous jumps and gaps in $p_s$ in these plots are not a result of numerical imprecision or algorithmic error. Finite number resolution restricts the number states of interest to a finite set, which is then further restricted to a smaller optimal set by limiting $p_e$ (see Subappendix~\ref{subsec:Probabilistic} for details). This set of interest typically exhibits continuous change in $p_s$ as $p_e$ and $r$ change, but at certain values of $r$ what constitutes the optimal set of number states of interest changes, resulting in discontinuous jumps of $p_s$.
	
	PSD operation gives us access to the previously discontinuous $p_s$ boost at $r=0.6585$ ($5.719$ dB of squeezing) from Ref.~\cite{64.3} discussed earlier, with very small $p_e$. This is an encouraging result, indicating that Zaidi's and van Loock's work on USD boosted BSM~\cite{64.3} is experimentally achievable for near-USD operation; i.e., PSD with very small permitted errors. When higher $p_e$ is allowed however, our analysis shows that the squeezing amount $r$ at which the maximum $p_s$ occurs is closer to that which maximizes $p_s$ for USD operation over the continuous regime (rather than where the discontinuous maximum occurs). In other words, $r=0.6585$~($5.719$ dB of squeezing) ceases to be the optimal squeezing as the allowed $p_e$ is raised. 
	\subsection{Measurement Confidence}
	\label{subsec:Measurement Confidence}
	
	Determining the effectiveness of a quantum error correction code to correct for an error introduced by a PSD BSM in some application, will require us to know the success (resp. failure) probability conditioned on the BSM circuit declaring a success. We call this conditional success probability the {\em measurement confidence}:
	\begin{equation}
	\alpha=\frac{p_s}{p_s+p_e}.
	\end{equation}
	
	With this in mind we take the data presented in Figure~\ref{fig:4 Rails, Squeezed, Prob, 7} and neglect $r$, instead analyzing~$p_s$ with respect to measurement confidence~($\alpha$), in Figure~\ref{fig:4 Rails, Squeezed, Prob, 7, Conf, Spec}. Here we see that each value of~$\alpha$ corresponds to a spectrum of points with varied~$p_s$~and~$p_e$, all with the same ratio~$p_s/(p_s + p_e)$ and therefore equivalent when viewed through the lens of quantum error correction, but with different success probabilities. 
	\begin{figure}[h!b!]
		\centering
		\includegraphics[width=\columnwidth]{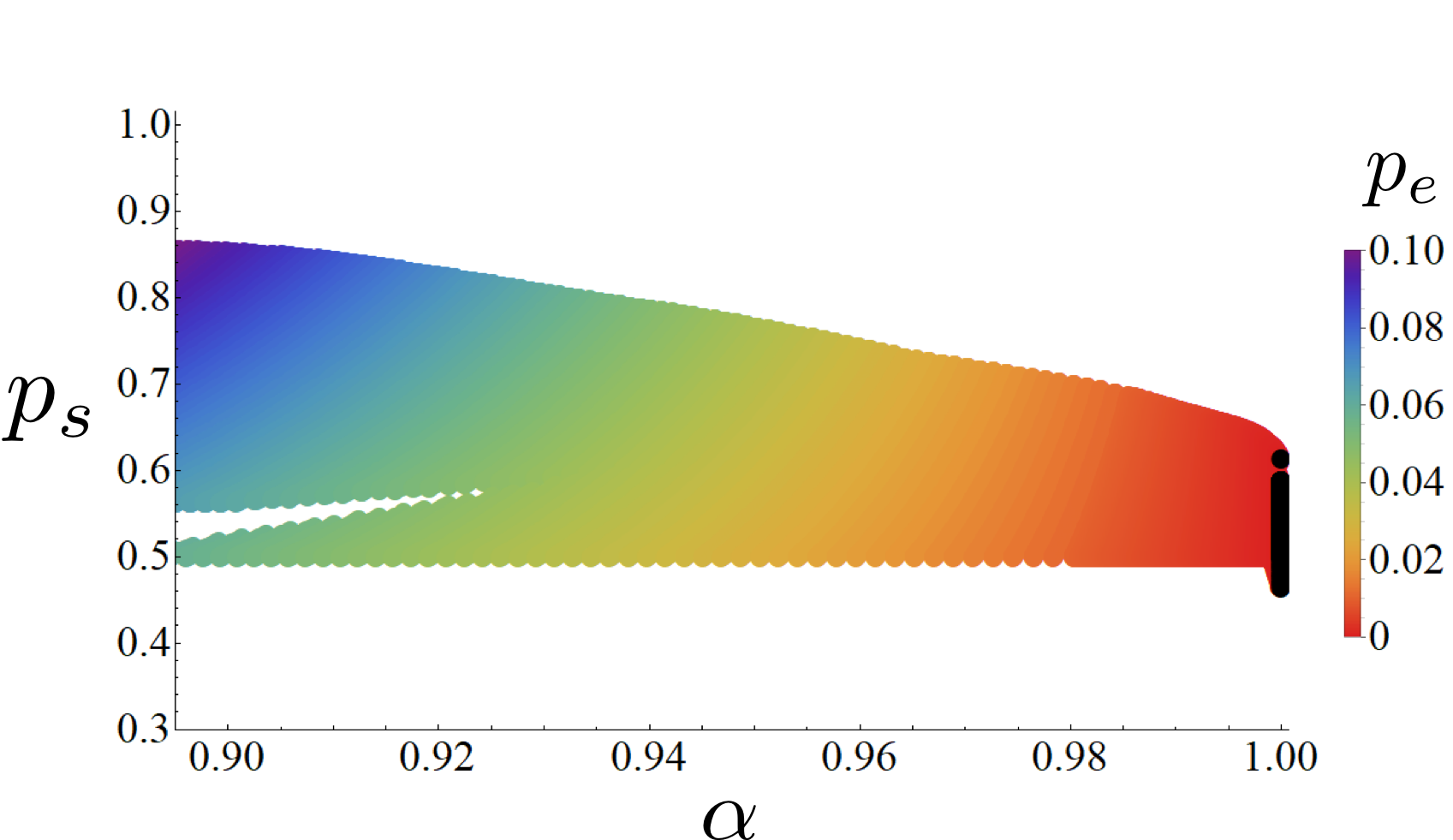}
		\caption{\label{fig:4 Rails, Squeezed, Prob, 7, Conf, Spec}Bell measurement success probability ($p_s$) vs. measurement confidence ($\alpha$), with respect to error probability ($p_e$), for a squeezed Bell measurement circuit, measured with PNR detectors restricted a PNR of $n_\text{max}=7$. The black dots represent the results of USD measurements with $p_e=0$.}
	\end{figure}
	
	From this plot we can obtain a curve showing the maximum success probability obtainable for a given measurement confidence. This curve is plotted in Figure~\ref{fig:4 Rails, Squeezed, Prob, Var, Conf, Env} for a variety of PNR limited detectors, including the current state-of-the-art limit of $n_\text{max}=7$. For ${n_\text{max}=7}$ we find a maximum $p_s$ of $85.8\%$ with ${\alpha=0.889}$, which is produced by $r=0.500$ ($4.343$ dB of squeezing). For ${n_\text{max}=\infty}$ we find a maximum $p_s$ of $\sim89.5\%$ with ${\alpha\approx0.90}$, which is produced by ${r=0.500}$~{($4.343$ dB of squeezing)}.
	\begin{figure}
		\centering
		\includegraphics[width=\columnwidth]{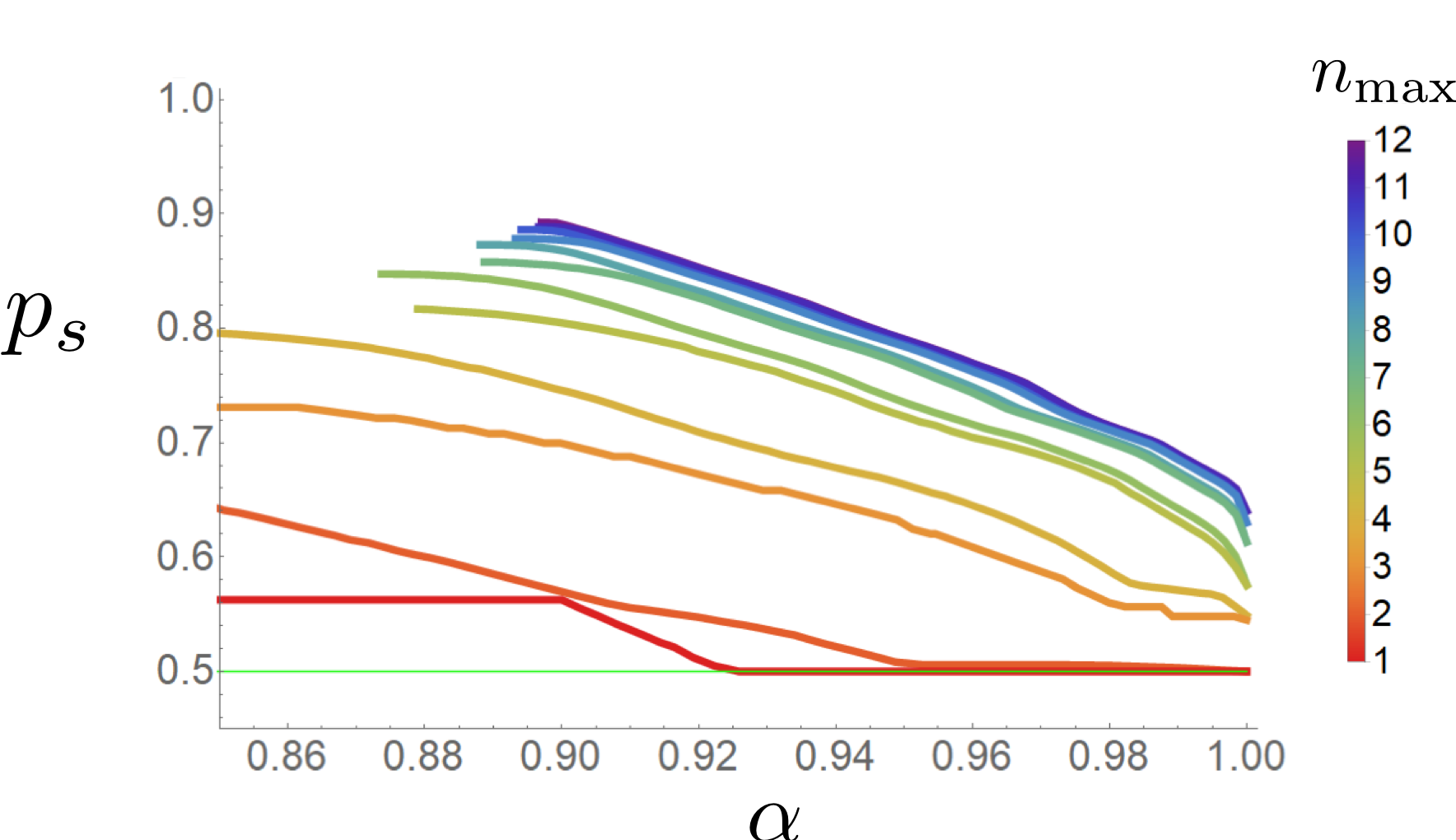}
		\caption{\label{fig:4 Rails, Squeezed, Prob, Var, Conf, Env}Maximum Bell measurement success probability ($p_s$) vs. measurement confidence ($\alpha$) for a squeezed Bell measurement circuit, shown for PNR detectors with PNR limited to $n_\text{max}\in[1,12]$. Each curve terminates at the maximum possible PSD $p_s$ which can be obtained for a given $n_\text{max}$.}
	\end{figure}
	\section{Detector Loss}
	\label{sec:Detector Loss}
	\begin{figure}[H]
		\centering
		\includegraphics[width=3in]{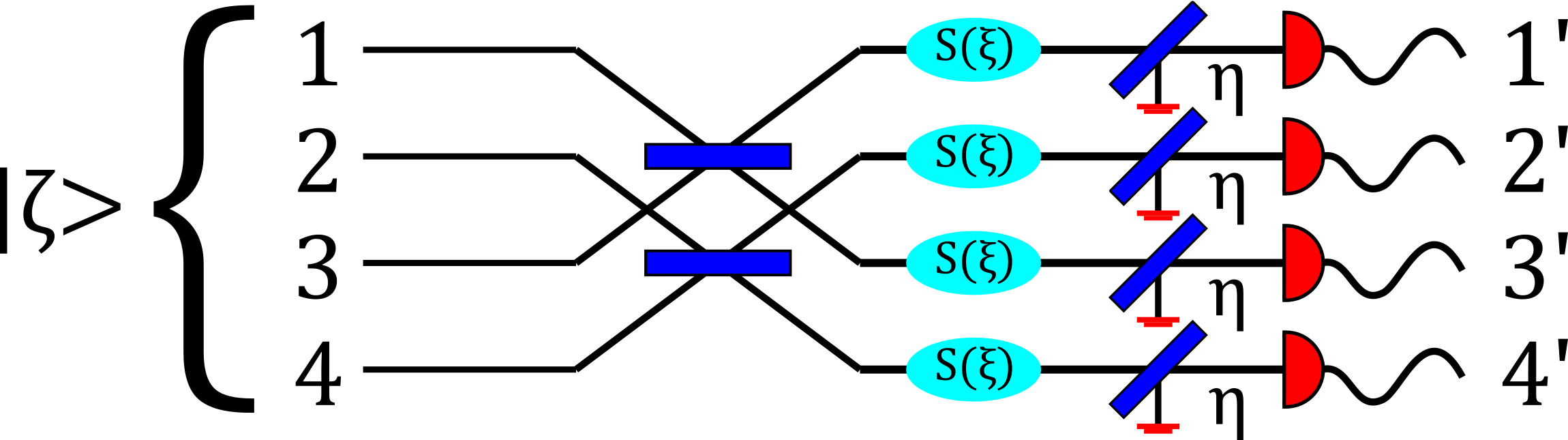}
		\caption{\label{fig:Lossy}Pre-detection quadrature squeezing Bell measurement setup with detector loss represented as pre-detection beam splitters.}
	\end{figure}
	
	Detector loss is an important impediment in any realistic optical system, especially in quantum information processing applications. Detector losses can be modeled as beam splitters with transmission~$\eta$ prior to an ideal detection, where the reflected photons in the beam splitters are `lost', and hence cannot be used to discriminate between number states. The process of calculating $p_s$ for the BSM circuit with lossy detectors, shown in Fig.~\ref{fig:Lossy}, is detailed in Subappendix~\ref{subsec:Lossy}.
	
	We assume the splitting ratios of all the beam splitters in the BSM circuit to be ideal (i.e., exactly 50-50), and furthermore assume all beam splitters to be lossless (i.e., total input photon number equals the total output photon number). This is because modern beam splitters can be engineered to have extremely low losses, and also because any losses in the beam splitters themselves can be subsumed into detector losses for the purposes of performance evaluation. 
	\begin{figure*}
		\centering
		\includegraphics[width=7in]{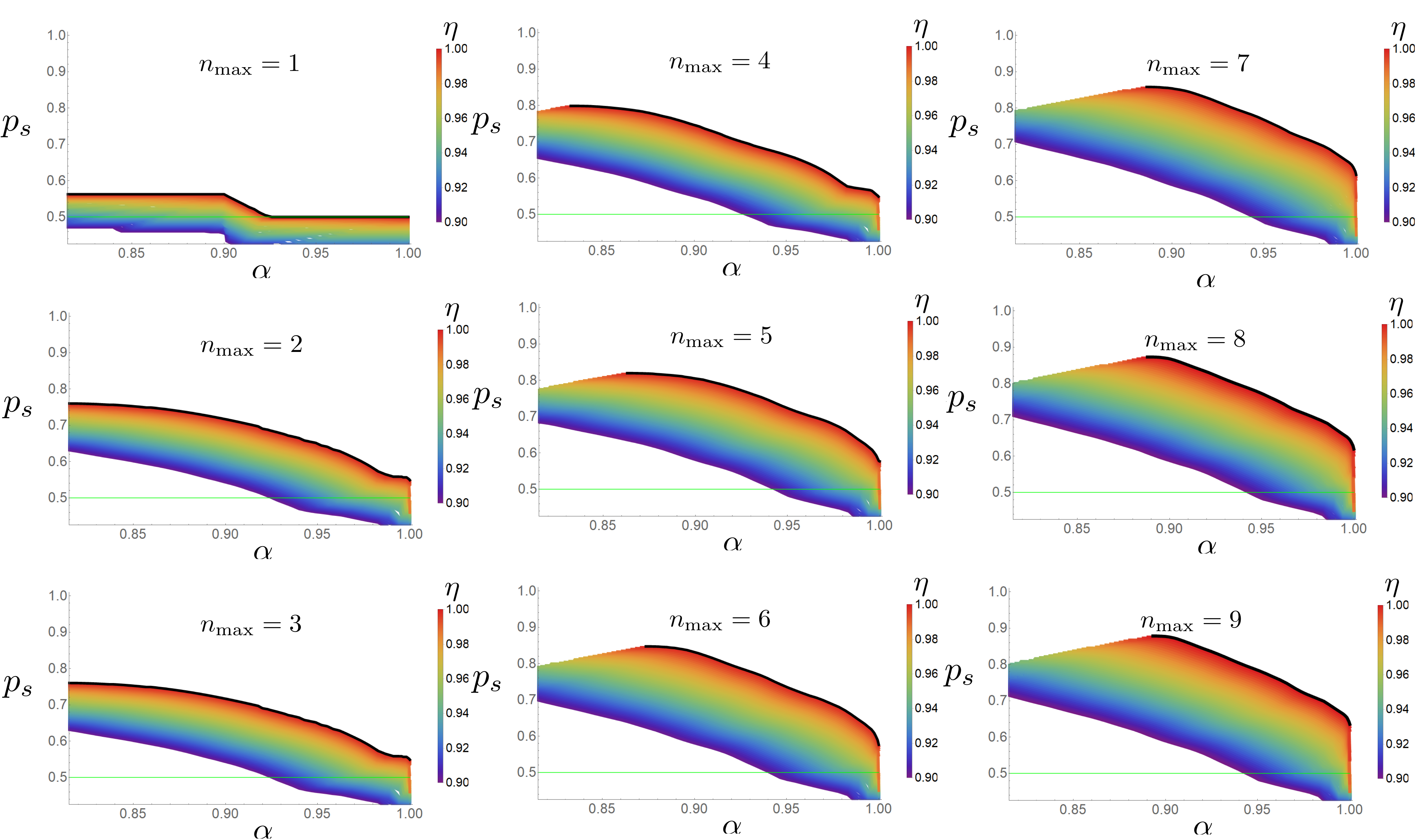}
		\caption{\label{fig:final}Maximum Bell measurement success probability ($p_s$) vs. measurement confidence ($\alpha$) for a squeezed Bell measurement circuit, shown for lossy PNR detectors with PNR limited to $n_\text{max}=[1,9]$ with efficiency $\eta\in[0.90,1]$. The black lines represent the success probabilities for lossless detectors shown in Figure~\ref{fig:4 Rails, Squeezed, Prob, Var, Conf, Env}.}
	\end{figure*}
	
	The ${n_\text{max}=7}$ transition edge sensing (TES) detectors previously cited were experimentally determined to have system device efficiencies ($\eta$) between 98\% and 100\%, which includes fiber coupling losses~\cite{TES}. With this value in mind, we evaluated the maximum success probability~($p_s$) obtainable for a given measurement confidence~($\alpha$), as in Figure~\ref{fig:4 Rails, Squeezed, Prob, Var, Conf, Env}, for $\eta\in[0.9,1]$ with PNR of ${n_\text{max}\in[1,9]}$, shown in Figure~\ref{fig:final}. For example, PNR detectors with ${n_\text{max}=7}$ and $\eta=0.98$ yield a maximum $p_s$ of $83.3\%$ with $\alpha=0.858$, which is achieved at $r=0.480$~{(4.169 dB of squeezing)}. 
	
	A full lossy analysis of squeezing-boosted BSMs with PNR $n_{\text{max}}\geq10$ could not be performed due to computational limitations. 
	\section{Discussion and conclusion}
	
	Injection of ancillary single photons or pre-detection quadrature squeezing can {\em boost} the success probability $p_s$ of linear optical Bell state measurement (BSM) to above the $50\%$ limit. The ability to achieve $p_s > 0.5$ has been found to be critical in resource-efficient realizations of linear optical quantum computing and all-photonic quantum repeaters~\cite{Percolation1}. Yet, neither of the aforesaid values of $p_s > 0.5$ are known to be the maximum achievable using single-photon ancillae and/or squeezing, thereby leaving it open whether $p_s \to 1$ might be achievable using these resources. An affirmative conclusion of this open question will not surprise us, given squeezing, linear optics, and photon number resolving (PNR) detection are in principle universal resources in quantum optics~\cite{CVuniversal}.
	
	We re-analyzed previous research predicting a maximum success probability, $p_s \approx 64.3\%$ for unambiguous-state discrimination (USD) based BSM on linear-optical qubits, employing pre-detection quadrature squeezing~\cite{64.3}. We showed that the above is a point result, and impossible to achieve experimentally since a small variation in the squeezing amplitude $r$ drops the USD success probability to about $59\%$. We showed that, rather unexpectedly, across a large range of squeezing, boosting the BSM success probability above the linear-optic limit of $50\%$ {\em is} possible while maintaining USD operation, but that this (experimentally achievable) maximum success rate is significantly lower, $59.6\%$, achievable with $r = 0.5774$ ($\sim 5$ dB of squeezing). 
	
	We then considered a probabilistic state discrimination (PSD) framework where we sacrifice pure USD operation, by allowing for a probability of error $p_e$ conditioned on a ``success" being declared (which happens with probability $p_s + p_e$). We find that with a small allowed $p_e$ the $64.3\%$ success probability result of Ref.~\cite{64.3} is achievable over a substantial range of~$r$. 
	
	For PSD measurements, we defined {\em measurement confidence} $\alpha = p_s/(p_s+p_e)$ as the probability of correct decision conditioned on a ``success" being declared. We found that PSD operation with significantly high error rates (e.g.,~${p_e \sim 0.1}$) can achieve success rates as high as $89.5\%$ with a measurement confidence of $\alpha \approx 0.89$, using ${\sim 5}$~dB of pre-detection squeezing, and PNR detectors with $7$~photon resolution. With photon number resolution and detector inefficiencies possible with the current state of the art transition edge sensor (TES) detectors~\cite{TES}, the maximum achievable success rate, $p_s \approx 83.3\%$ with a measurement confidence, $\alpha \approx 0.858$. 
	
	It has recently been shown that arbitrarily large cluster states for all-photonic universal quantum computation can be produced with a supply of $3$-photon GHZ states, and a Bell state measurement device with success probability $p_s$ exceeding a threshold of $\approx 0.59$ (with USD operation). This used ideas from percolation theory~\cite{Percolation1}. Detector losses within the usual Bell-state measurement (BSM) circuits are considered heralded losses since they can be subsumed within $p_s$, and maintain USD operation. A complete analysis of percolation-based direct generation of universal photonic cluster states with unheralded losses (for example, photon losses {\em after} the creation of the cluster) has not yet been done, leaving open an important question of whether fault tolerant cluster state quantum computing with ballistic cluster creation is possible with probabilistic boosted BSMs {\em and} unheralded photon losses. Fault tolerant universal cluster model quantum computing is known to be possible with single qubit measurements with conditional probabilities of error below $0.75\%$ (i.e., measurement confidence above $0.9925$), where the error model includes preparation, gate, storage and measurement errors~\cite{Percolation2}.
	
\begin{figure}[!h!tp]
		\centering
		\includegraphics[width=3in]{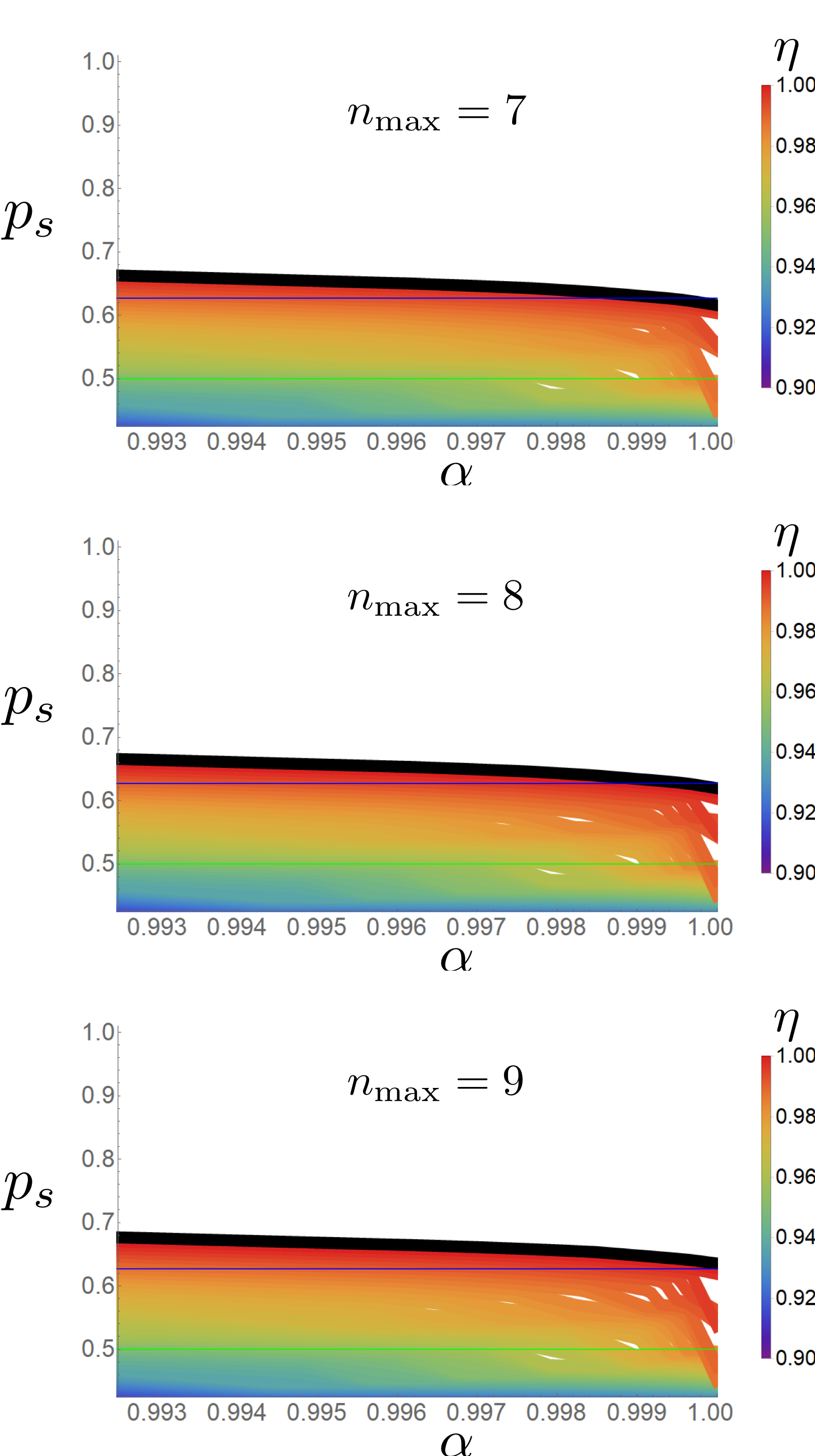}
		\caption{\label{fig:finalShort}Maximum Bell measurement success probability ($p_s$) vs. measurement confidence ($\alpha$) for a squeezed Bell measurement circuit, shown for lossy PNR detectors with PNR limited to $n_\text{max}=[7,9]$ with efficiency $\eta\in[0.90,1]$.}
	\end{figure}

	Our PSD analysis of boosted BSMs can form a starting point of an analysis to see if there is an operational point in the $p_s$-vs.-$p_e$ trade-space where both the percolation condition for long-range connectivity on a subgraph of the Rausendorf-Harrington lattice {\em and} the error threshold for single-qubit measurements for fault tolerant operation are satisfied simultaneously. We leave this open for future work. The gate errors in the Rausendorf-Harrington construction for fault tolerant cluster state quantum computing does not quite translate to BSM error rates (the latter translates to a control-phase two-qubit measurement). However, just to serve as an illustration, in Figure~\ref{fig:finalShort} we show the maximum success probabilities obtainable within a measurement confidence of $0.9925$ and above. From this figure it appears that squeezing-boosted BSM devices and PNR detectors with $n_\text{max}\geq7$ might suffice to realize all-photonic universal quantum computation, with $3$-photon GHZ states as the initial resource. 
	
	Similarly, we believe that our PSD analysis of boosted BSMs can form a starting point of an analysis of a rate-vs.-distance calculation for quantum repeaters based on entanglement sources, mode multiplexing and BSMs~\cite{MultiplexedRepeater2015}. The objective of such an analysis would be to see if there is an optimal operational point in the $p_s$-vs.-$p_e$ trade-space where the rate-vs.-loss exponent is minimized, leading to the best possible entanglement generation rate at a given distance, when in-line squeezing is used within the BSMs. We leave this for future work as well.
	\newpage{\pagestyle{empty}}
	\acknowledgements
	
	This research was funded by an NSF subaward of a Yale University led project, grant number 1640959, ``EFRI ACQUIRE: Integrated nanophotonic solid state memories 
	for telecom wavelength quantum repeaters".

	\bibliography{bib}
	
	\newpage{\pagestyle{empty}\cleardoublepage}
	\appendix
	
	\section{Numerically Computing Bell Measurement Success Probabilities}
	\label{ap:Numerical}
	
	\subsection{State Construction}
	\label{subsec:State Construction}
	
	The first step in evaluating the success probability of a Bell measurement system is constructing the modified Bell states the system outputs onto its detectors. To do so we used a simple replacement algorithm for creation operators, derived from~(\ref{eq:Beam Splitter}), where the action of a beamsplitter on modes $j\&k$ is performed as shown in~(\ref{eq:Beamsplitter Substitution}).
	\begin{subequations}
		\label{eq:Beamsplitter Substitution}
		\begin{eqnarray}
		a_j^{\dag}&\rightarrow& \frac{1}{\sqrt{2}}\left(i {a^\prime}_j^{\dag}+{a^\prime}_k^{\dag}\right)
		\\
		a_k^{\dag}&\rightarrow& \frac{1}{\sqrt{2}}\left({a^\prime}_j^{\dag}+i {a^\prime}_k^{\dag}\right)
		\end{eqnarray}
	\end{subequations}
	
	For a system which involves quadrature squeezing, this replacement algorithm is complicated by the infinite quantity of number states produced by a squeezing operation. When numerically constructing a squeezed state we must used a truncated version of the squeezing operation~(\ref{eq:Squeezed Number States}) which does not make use of an infinite series. We show such a truncated squeezing operation below in~(\ref{eq:Truncated Squeezed Number States}), where the infinite series is truncated by the maximum number state $|k_{\text{max}}\rangle$.
	\begin{subequations}
		\label{eq:Truncated Squeezed Number States}
		\begin{eqnarray}
		S(\xi)\left(a_i^{\dag}\right)^n &\rightarrow& \left(\frac{1}{\text{cosh}~2r}\right)^{n+1/2}n!
		\\
		&&\sum_{j=0}^{\left[\frac{n}{2}\right]}\frac{\left(-\xi^*\right)^j\left(\text{cosh}~2r\right)^{2j}}{\left(n-2j\right)!j!} 
		\nonumber \\
		&&\sum_{k=0}^{k_{\text{max}}}\frac{\xi^k}{k!}\left({a^\prime}_i^{\dag}\right)^{n-2j+2k}
		\nonumber \\
		\xi &\equiv& \frac{1}{2}e^{i\phi}\text{tanh}~r
		\\
		\text{dB} &=& -10 \log_{10}\left(e^{-2r}\right)
		\end{eqnarray}
	\end{subequations}
	
	Systems with finite PNR $n_{\text{max}}$ can be exactly evaluated, as truncation is already implied by their finite PNR detectors. Setting $k_{\text{max}}=\frac{1}{2}\left(n_{\text{max}}-n+2j\right)$ ensures that all of the states within the detection threshold, and only those states, will be evaluated. (Note that $j$ is the summation index of the outer sum within which $\sum_{k}$ is contained).
	
	To approximate arbitrary photon number discrimination we set $n_{\text{max}}$ such that the probability sum of each modified Bell state exceeds $0.999$ for $r\leq0.70$ and $0.995$ for $0.70<r\leq0.90$. This threshold can be achieved by $n_{\text{max}}=13$. In our calculations $n_{\text{max}}=13$ is used to model $n_{\text{max}}=\infty$.
	
	(Note that as $r$ increases more probability gets shifted into larger number states, requiring a higher value of $n_{\text{max}}$ to fully capture the effects of squeezing).
	
	With this we can produce a finite list of number states and their amplitudes for each modified Bell state, as shown in Table~\ref{tab:example}, where each Bell state is scaled by the square root of its relative probability. (For an equal distribution of Bell states, each state has a $\frac{1}{4}$ probability of occurrence). 
	\begin{table*}[t!]
		\caption{\label{tab:example}List of scaled number states from system detailed in Section~\ref{sec:squeezing}, Figure~\ref{fig:Squeezed}}
		\begin{ruledtabular}
			\begin{tabular}{r|r|r|r}
				$\frac{1}{2}|\psi^{\prime\prime}_+(r)\rangle$ & $\frac{1}{2}|\psi^{\prime\prime}_-(r)\rangle$ & $\frac{1}{2}|\phi^{\prime\prime}_+(r)\rangle$ & $\frac{1}{2}|\phi^{\prime\prime}_-(r)\rangle$ 
				\\ \hline
				$\frac{i\left(\text{sech}~r\right)^4}{2\sqrt{2}}|1100\rangle$ & $\frac{\left(\text{sech}~r\right)^4}{2\sqrt{2}}|0110\rangle$ & $-\frac{i}{\sqrt{2}}\left(\text{sech}~r\right)^2\text{tanh}~r|0000\rangle$ & $\frac{i\left(\text{sech}~r\right)^4}{4}|2000\rangle$ 
				\\
				$\frac{i\left(\text{sech}~r\right)^4}{2\sqrt{2}}|0011\rangle$ & $-\frac{\left(\text{sech}~r\right)^4}{2\sqrt{2}}|1001\rangle$ &$-\frac{i\left(\text{sech}~r\right)^4\left(\text{cosh}~2r-2\right)}{4}|2000\rangle$ & $-\frac{i\left(\text{sech}~r\right)^4}{4}|0200\rangle$ 
				\\
				$\frac{i\left(\text{sech}~r\right)^4\text{tanh}~r}{4}|0211\rangle$ & $\frac{\left(\text{sech}~r\right)^4\text{tanh}~r}{4}|0112\rangle$ & $-\frac{i\left(\text{sech}~r\right)^4\left(\text{cosh}~2r-2\right)}{4}|0200\rangle$ & $\frac{i\left(\text{sech}~r\right)^4}{4}|0020\rangle$ 
				\\
				$\frac{i\left(\text{sech}~r\right)^4\text{tanh}~r}{4}|2011\rangle$ & $\frac{\left(\text{sech}~r\right)^4\text{tanh}~r}{4}|2110\rangle$ & $-\frac{i\left(\text{sech}~r\right)^4\left(\text{cosh}~2r-2\right)}{4}|0020\rangle$ & $-\frac{i\left(\text{sech}~r\right)^4}{4}|0002\rangle$ 
				\\
				\vdots & \vdots & \vdots & \vdots
			\end{tabular}
		\end{ruledtabular}
	\end{table*}

	\subsection{Separating Number States}
	\label{subsec:Separating Number States}
	
	Now that we have a finite list of number states for each modified Bell state, the next step is to arrange the number states by multiplicity with which they occur in parent Bell states, as shown in Table~\ref{tab:separated_example}. Number states which occur in only one Bell state go to a unique number states list. Number states which occur in multiple Bell states go to a duplicate number states list. (Note: this is the most computationally intensive step of computing Bell measurement efficiencies.)
	\begin{table*}[t!]
		\caption{\label{tab:separated_example}List of scaled number states from system detailed in Section~\ref{sec:squeezing}, Figure~\ref{fig:Squeezed}, separated by uniqueness}
		\begin{ruledtabular}
			\begin{tabular}{r|r}
				Unique Number States List $|n_1,n_2,n_3,n_4\rangle_\xi$ & Duplicate Number States List $|n_1,n_2,n_3,n_4\rangle_\xi$
				\\ \hline
				$-\frac{i}{\sqrt{2}}\left(\text{sech}~r\right)^2\text{tanh}~r|0000\rangle_{\phi_+}$ & $-\frac{i\left(\text{sech}~r\right)^4\left(\text{cosh}~2r-2\right)}{2}|2000\rangle_{\phi_+}~~,~~~\frac{i\left(\text{sech}~r\right)^4}{4}|2000\rangle_{\phi_-}$ 
				\\
				$\frac{i\left(\text{sech}~r\right)^4}{2\sqrt{2}}|1100\rangle_{\psi_+}$ & $-\frac{i\left(\text{sech}~r\right)^4\left(\text{cosh}~2r-2\right)}{2}|0200\rangle_{\phi_+}~~,-\frac{i\left(\text{sech}~r\right)^4}{4}|0200\rangle_{\phi_-}$
				\\
				$\frac{\left(\text{sech}~r\right)^4}{2\sqrt{2}}|0110\rangle_{\psi_-}$ & $-\frac{i\left(\text{sech}~r\right)^4\left(\text{cosh}~2r-2\right)}{2}|0020\rangle_{\phi_+}~~,~~~\frac{i\left(\text{sech}~r\right)^4}{4}|0020\rangle_{\phi_-}$
				\\
				$-\frac{\left(\text{sech}~r\right)^4}{2\sqrt{2}}|1001\rangle_{\psi_-}$ & $-\frac{i\left(\text{sech}~r\right)^4\left(\text{cosh}~2r-2\right)}{2}|0002\rangle_{\phi_+}~~,-\frac{i\left(\text{sech}~r\right)^4}{4}|0002\rangle_{\phi_-}$
				\\
				\vdots & \vdots
			\end{tabular}
		\end{ruledtabular}
	\end{table*}

	\subsection{USD Measurements}
	\label{subsec:USD}
	
	To find the USD success probability of a Bell measurement system all one has to do is sum together the modulus squares of all the coefficient terms in the unique number states list.

	\subsection{PSD Measurements}
	\label{subsec:Probabilistic}
	
	To begin an analysis of PSD success probability we must first note two things. First, not all duplicate number states can provide a boost in Bell state discrimination. Number states which have equal probability in all their parent Bell states have no marginal distinguishability and must be excluded from our analysis. Second, determination of which Bell state is the most probable parent state for a given number state can vary with changes to system parameters. Therefore any state selection rule applied to number states must be applied on a case by case basis for all parameter values being examined. 
	
	To identify an optimal family of duplicate number states for a given error probability and for a given set of parameters we first numerically evaluate the duplicate number state amplitudes corresponding to each Bell state for the given parameters. We then transform the scaled duplicate number state amplitudes to probabilities. Next we identify the total probability for each duplicate number state, over all parent Bell states, and the largest probability provided by an individual Bell state. These two values can be used to compute the expected $p_s$ and $p_e$ increases for making Bell state decisions based on a given duplicate number state. 
	
	These success probability and error probability increases are then stored as pairs in a list. We then order the list by the $p_s$ to $p_e$ ratios of each number state, greatest to least. Each number state is then evaluated in order according to a simple algorithm. If its $p_e$ increase plus the $p_e$ increases of all other number states thus far included is less than the maximum allowable $p_e$, the number state is included and its efficiency added to the system's total efficiency. If not, it is excluded. All duplicate number states are evaluated in this manner. 
	
	(Note that this algorithm only provides an approximately optimal family of duplicate number states under the condition that there exists a very large number of duplicate number states with small $p_e$ and few if any duplicate number states have high $p_e$. This condition is satisfied for all systems analyzed in this paper, but cannot be guaranteed for all systems in general).

	\subsection{Lossy Measurements}
	\label{subsec:Lossy}
	
	Lossy states are constructed exactly as above in Subappendix~\ref{subsec:State Construction}, using~(\ref{eq:Loss_App}) as the replacement algorithm to model loss. 
	\begin{equation}
	a_{i}^{\dag} \rightarrow \sqrt{\eta}~{a^\prime}_{i}^{\dag} + \sqrt{1-\eta}~l_{i}^{\dag}
	\label{eq:Loss_App}
	\end{equation}
	
	However it must be noted that the introduction of loss means that setting $k_{\text{max}}=\frac{1}{2}(n_{\text{max}}-n+2j)$ no longer ensures all states within the detection threshold are evaluated. Doing so excludes larger number states ${\exists~i~\text{s.t}.~n_i>k_{\text{max}}}$ which due to loss are detected as number states within the detection threshold ${\forall~i, n_i\leq k_{\text{max}}}$. We compensate for this by setting $n_{\text{max}}$ two higher than its actual value during the state construction algorithm, including the application of loss, and then deleting all creation operators with powers greater than $n_{\text{max}}$ before evaluating probabilities. 
	
	Additionally, each Bell state may now have multiple number states which can be detected as the same photon number pattern, for instance $|1001\rangle$ and $l_2^{\dag}l_3^{\dag}|1001\rangle$ are both detected as ${N_1N_2N_3N_4=1001}$. Probabilities for photon number patterns are found by summing the modulus squares of all number states with identical detector patterns within a modified Bell state.

	\section{Non-Uniform and Phase-Shifted Squeezing}
	\label{ap:Non-Uniform}
	
	Any numerical analysis of non-uniform quadrature squeezing on Bell measurement systems, where it is not true that ${r_1=r_2=r_3=r_4}$, is complicated by the exponential order of such calculations. With four separately squeezed modes, a coarse scan over squeezing intensity ${r_i\in[0,0.85]}$ with twenty values fore each $r_i$ requires over one hundred thousand simulation data. This is beyond the scope of the computational capabilities our research group has at its disposal, and we therefore cannot draw any firm conclusions about the potential effects of asymmetric mode squeezing. 
	
	However, we did perform a very coarse scan of ${r_i\in\left\{{0,0.15,0.30,0.45,0.60,0.6585,0.75}\right\}~\forall~i}$, with ${\theta_i=0~\forall~i}$. For the 2401 such points analyzed, there existed no datum with non-uniform squeezing parameters which outperformed all uniformly squeezed data with a shared squeezing parameter. With this in mind we concluded that is unlikely for non-uniform squeezing intensities to produce improvements over uniform squeezing intensities, but we cannot provide conclusive numerical evidence to back up this assertion. 
	
	Non-uniform squeezing phases present an even more computationally intractable problem. Performing enough computations to make even vague assumptions about non-uniform squeezing phases proved too difficult to attempt, and we restricted ourselves to ${\theta_i=0~\forall~i}$ to keep the computational complexity of the problem manageable. We have no particular reason to believe ${\theta_i=0~\forall~i}$ is optimal. Exploring non-uniform squeezing phases may provide improvements over our results and could be a fruitful area of research for a research group looking to analyze similar systems, particularly with analytic approaches which may be less limited by the computationally intractable nature of the problem.

\end{document}